\documentclass[dvips]{acta}
\usepackage{supertabular,lscape,epsfig}
\usepackage{amssymb}
\usepackage{amsmath}
\DeclareSymbolFont{ppa}{OT1}{ppl}{m}{it}
\DeclareMathSymbol{\vv}{\mathalpha}{ppa}{'166}

\newfont{\hb}{rphvb at 10pt}
\newfont{\hbo}{rphvbo at 10pt}
\newfont{\bitt}{rptmbi at 12pt}
\newfont{\bits}{rptmbi at 11pt}

\SetPages{251}{265}

\SetVol{65}{2015}

\begin{document}

\newcommand{\TabApp}[2]{\begin{center}\parbox[t]{#1}{\centerline{
  {\bf Appendix}}
  \vskip2mm
  \centerline{\small {\spaceskip 2pt plus 1pt minus 1pt T a b l e}
  \refstepcounter{table}\thetable}
  \vskip2mm
  \centerline{\footnotesize #2}}
  \vskip3mm
\end{center}}

\newcommand{\TabCapp}[2]{\begin{center}\parbox[t]{#1}{\centerline{
  \small {\spaceskip 2pt plus 1pt minus 1pt T a b l e}
  \refstepcounter{table}\thetable}
  \vskip2mm
  \centerline{\footnotesize #2}}
  \vskip3mm
\end{center}}

\newcommand{\TTabCap}[3]{\begin{center}\parbox[t]{#1}{\centerline{
  \small {\spaceskip 2pt plus 1pt minus 1pt T a b l e}
  \refstepcounter{table}\thetable}
  \vskip2mm
  \centerline{\footnotesize #2}
  \centerline{\footnotesize #3}}
  \vskip1mm
\end{center}}

\newcommand{\MakeTableApp}[4]{\begin{table}[p]\TabApp{#2}{#3}
  \begin{center} \TableFont \begin{tabular}{#1} #4 
  \end{tabular}\end{center}\end{table}}

\newcommand{\MakeTableSepp}[4]{\begin{table}[p]\TabCapp{#2}{#3}
  \begin{center} \TableFont \begin{tabular}{#1} #4 
  \end{tabular}\end{center}\end{table}}

\newcommand{\MakeTableee}[4]{\begin{table}[htb]\TabCapp{#2}{#3}
  \begin{center} \TableFont \begin{tabular}{#1} #4
  \end{tabular}\end{center}\end{table}}

\newcommand{\MakeTablee}[5]{\begin{table}[htb]\TTabCap{#2}{#3}{#4}
  \begin{center} \TableFont \begin{tabular}{#1} #5 
  \end{tabular}\end{center}\end{table}}

\newfont{\bb}{ptmbi8t at 12pt}
\newfont{\bbb}{cmbxti10}
\newfont{\bbbb}{cmbxti10 at 9pt}
\newcommand{\uprule}{\rule{0pt}{2.5ex}}
\newcommand{\douprule}{\rule[-2ex]{0pt}{4.5ex}}
\newcommand{\dorule}{\rule[-2ex]{0pt}{2ex}}
\begin{Titlepage}
\Title{Empirical Conversions of Broad-Band Optical and Infrared Magnitudes to Monochromatic Continuum Luminosities\\for Active Galactic Nuclei}
\Author{S.~~K~o~z~³~o~w~s~k~i}{Warsaw University Observatory, Al. Ujazdowskie 4, 00-478 Warszawa, Poland\\
e-mail: simkoz@astrouw.edu.pl}

\Received{August 15, 2015}
\end{Titlepage}

\Abstract{We use public data for 105\,783 quasars from The Sloan Digital
  Sky Survey (SDSS) Data Release~7 (DR7) that include spectral
  monochromatic luminosities at 5100~\AA, 3000~\AA, and 1350~\AA, and the
  corresponding observed broad-band {\it ugriz, VRI} (converted), {\it JHK}
  and WISE magnitudes, and derive broad-band--to--monochromatic luminosity
  ratios independent of a cosmological model. The ratios span the redshift
  range of $z=0.1\div4.9$ and may serve as a proxy for measuring the
  bolometric luminosity, broad line region (BLR) radii and/or black hole
  masses, whenever flux-calibrated spectra are unavailable or the existing
  spectra have low signal-to-noise ratios. They are provided both in
  tabular and parametric form.}{Galaxies: active -- quasars: general --
  Techniques: photometric}

\Section{Introduction}
Active galactic nuclei (AGNs) owe their tremendous brightness to accretion
disks forming around supermassive black holes at their centers. It is
presumed that the disk temperature $T$ falls off with the disk radius $R$
as $T\propto R^{-3/4}$ (\eg Shakura and Sunyaev 1973), giving rise to a
wide range of photon energies at a continuous range of wavelengths -- a
feature known as the ``continuum'' in observed AGN spectra. This continuum
can be described either at particular wavelengths (monochromatic
luminosities) or in standard broad-band filters (broad-band luminosities).

Continuum photons, either monochromatic, broad-band, or bolometric,
may provide key diagnostics in understanding AGN physics, as they
respond to the hard UV ionizing photons. A fraction of these UV
photons is partially absorbed by the gas-dust clouds away from the
disk, in the broad line region (BLR), and re-emitted in a form of
broad emission lines at wavelengths corresponding to certain
differences between energy levels in atoms and molecules. The typical
distance to the clouds is $r=c\tau$, so they reverberate any
luminosity changes with time-lags $\tau$ that are strongly correlated
with the continuum luminosity $L$ as $\tau\approx r \propto
L^{1/2}$. This is commonly known as the BLR radius--luminosity
relation (\eg Kaspi \etal 2000, 2007, Bentz \etal 2009). The
implications of this relation are essential in the determination of
the central black hole mass $M_{BH}$ {\it via} the viral theorem
$M_{BH}\propto L^{1/2}\vv^2$, where $\vv$ is the velocity of the BLR
gas-dust clouds (\eg Vestergaard and Peterson 2006). Their velocity is
routinely measured from AGN spectra as either full width (of the broad
emission line) at its half maximum (FWHM) or dispersion~($\sigma$).

To measure the delay between the continuum and line luminosity variations,
it is a common practice in reverberation mapping studies to measure the
monochromatic continuum fluxes in the vicinity of the reverberating lines.
For the $H\beta$ line it is the luminosity at 5100~\AA, for the MgII line at
3000~\AA, and for the CIV line at 1350~\AA. Knowing these luminosities and a
BLR radius--luminosity relation (Kaspi \etal 2000, 2007, Bentz \etal 2009),
one can estimate the expected BLR radius and hence the central black hole
mass.

Monochromatic luminosities can be turned into estimated time-lags for these
lines, and may, for example, help in designing a spectroscopic experiment
to observe them {\it via} spectroscopic (\eg Peterson 1993, Denney \etal
2010, Shen \etal 2015) or photometric (\eg Chelouche and Daniel 2012, Zu
\etal 2013b) reverberation mapping. Inverting the problem, once a time-lag
between a continuum flux and a reverberating line flux is measured, the
absolute luminosity of an AGN is known, making it a ``standardizable
candle'' (\eg Watson \etal 2011, Czerny \etal 2013).

Our main motivation here is to estimate the empirical monochromatic
luminosities at these three rest-frame wavelengths (5100~\AA, 3000~\AA,
1350~\AA) from the broad-band AGN magnitudes, when we lack flux-calibrated
spectra or they are too noisy. We simply take the $\approx100\,000$ AGNs with
spectroscopically measured monochromatic luminosities, black hole masses,
emission lines, spectral slopes (Shen \etal 2011) and estimate the
monochromatic fluxes based on the wealth of broad-band optical--IR
magnitudes that are available (Schneider \etal 2010).
\vspace*{-9pt}
\Section{Data}
\vspace*{-5pt}
To measure the monochromatic fluxes from the broad-band filters, we have
downloaded the data for 105\,783 quasars from Data Release 7 of SDSS from
Schneider \etal (2010) and Shen \etal (2011). All of them are brighter than
$M_i<-22$~mag and have at least one broad emission line with a FWHM larger
than 1000~km/s or interesting absorption features. Schneider \etal (2010)
provide a dataset containing both the observed SDSS {\it ugriz} AB
magnitudes for these objects as well as matched {\it JHK} and WISE
magnitudes. We corrected these observed magnitudes for Galactic extinction
using the extinction maps of Schlegel, Finkbeiner and Davis (1998).
Schneider \etal (2010) already provide $u$-band extinction ($A_u$) for all
sources and we convert $A_u$ to other wavelengths using the $R_V=3.1$
Galactic extinction curve (Cardelli, Clayton and Mathis 1989) with the
following values: $A_g/A_u=0.736$, $A_r/A_u=0.534$, $A_i/A_u=0.405$,
$A_z/A_u=0.287$, $A_U/A_u=1.052$, $A_V/A_u=0.641$ (unused), $A_R/A_u=0.520$
(unused), $A_I/A_u=0.373$ (unused), $A_J/A_u=0.176$, $A_H/A_u=0.111$,
$A_K/A_u=0.072$, $A_{W1}/A_u=0.033$, $A_{W2}/A_u=0.016$,
$A_{W3}/A_u=0.000$, and $A_{W4}/A_u=0.000$. The extinctions in {\it V},
{\it R}, and {\it I} bands were not used as these magnitudes were
synthesized directly from the extinction-corrected {\it ugriz} magnitudes.

These quasars generally do not have the common Johnson-Cousins {\it VRI}
magnitudes and we derive them directly from the extinction-corrected {\it
  ugriz} magnitudes. First, we calculate the $r-R$, $r-I$, $i-R$ and $i-I$
synthetic colors as a function of redshift using an average AGN spectrum
from Richards \etal (2006a) and the respective filter transmission
curves. We then match the SDSS dataset to the 9~deg$^2$ AGN and Galaxy
Evolution Survey (AGES; 152 AGN matches; Kochanek \etal 2012), containing
{\it R} and {\it I} magnitudes, and shift the synthetic colors such that
the converted {\it R} and {\it I} magnitudes match the ones observed in the
NOAO Deep, Wide-Field Survey (NDWFS; Jannuzi and Dey 1999). We create {\it
  R} and {\it I} magnitudes from two sets of colors to verify their
correctness (the dispersion is 0.12~mag).

\MakeTableSepp{l@{\hspace{1.5cm}}r@{\hspace{1cm}}r@{\hspace{1.5cm}}r@{\hspace{1cm}}r}{12.5cm}
{Basic properties of common astronomical filters}
{\hline 
\uprule
Filter & Central            & flux at $m=0$ & Frequency & Ref.\\
\dorule
       & $\lambda$ [$\mu$m] & [Jy]          & [Hz]      & \\
\hline
\noalign{\vskip7pt}
\multicolumn{5}{c}{Galex}\\
\hline
\uprule
FUV (Vega) & 0.1539 & 514.98 & $1.948\times10^{15}$ & (1) \\
FUV (AB)   & 0.1539 & 3631.0 & $1.948\times10^{15}$ & (1)  \\
NUV (Vega) & 0.2316 & 781.84 & $1.295\times10^{15}$ & (1)  \\
NUV (AB)   & 0.2316 & 3631.0 & $1.295\times10^{15}$ & (1)  \\
\hline
\noalign{\vskip7pt}
\multicolumn{5}{c}{{\it UBVRI} (Vega)}\\
\hline
\uprule
U  & 0.375 & 1823.0  & $7.994\times10^{14}$ & (2)  \\
B  & 0.430 & 4130.0  & $6.972\times10^{14}$ & (2)  \\
V  & 0.554 & 3636.0  & $5.414\times10^{14}$ & (2)  \\
R  & 0.641 & 3080.0  & $4.677\times10^{14}$ & (2)  \\
I  & 0.789 & 2416.0  & $3.802\times10^{14}$ & (2)  \\
\hline
\noalign{\vskip7pt}
\multicolumn{5}{c}{SDSS (AB)}\\
\hline
\uprule
u  & 0.3543 &  3631.0 & $8.462\times10^{14}$ & (3)  \\
g  & 0.4770 &  3631.0 & $6.285\times10^{14}$ & (3)  \\
r  & 0.6231 &  3631.0 & $4.811\times10^{14}$ & (3)  \\
i  & 0.7625 &  3631.0 & $3.932\times10^{14}$ & (3)  \\
z  & 0.9134 &  3631.0 & $3.282\times10^{14}$ & (3) \\
\hline
\noalign{\vskip7pt}
\multicolumn{5}{c}{2MASS (Vega)}\\
\hline
\uprule
J  &  1.235 & 1594.0  & $2.427\times10^{14}$ & (4)  \\
H  &  1.662 & 1024.0  & $1.804\times10^{14}$ & (4)  \\
K  &  2.159 &  666.7  & $1.389\times10^{14}$ & (4)  \\
\hline
\noalign{\vskip7pt}
\multicolumn{5}{c}{Spitzer (Vega)}\\
\hline
\uprule
$[3.6]$ &  3.561 &  280.9  & $8.419\times10^{13}$ & (5)  \\
$[4.5]$ &  4.509 &  179.7  & $6.649\times10^{13}$ & (5)  \\
$[5.8]$ &  5.693 &  115.0  & $5.266\times10^{13}$ & (5)  \\
$[8.0]$ &  7.982 &   64.13 & $3.756\times10^{13}$ & (5)  \\
$[24]$  & 23.68  &    7.17 & $1.265\times10^{13}$ & (6)  \\
$[70]$  & 71.42  &   0.778 & $4.198\times10^{12}$ & (7)  \\
$[160]$ & 155.9  &   0.160 & $1.923\times10^{12}$ & (8)  \\
\hline
\noalign{\vskip7pt}
\multicolumn{5}{c}{WISE (Vega)}\\
\hline
\uprule
W1 &  3.353 & 309.540 & $8.850\times10^{13}$ & (9)  \\
W2 &  4.603 & 171.787 & $6.445\times10^{13}$ & (9)  \\
W3 & 11.561 &  31.674 & $2.675\times10^{13}$ & (9)  \\
W4 & 22.088 &   8.363 & $1.346\times10^{13}$ & (9) \\
\hline
\noalign{\vskip5pt}
\multicolumn{5}{p{12cm}}{In Column~2 the central wavelength of a filter is given,
in Column~3 the flux [Jy] for an object of zero magnitude is given, and in Column~4 the corresponding frequency
to the central wavelength of a filter is provided. References in Column~5 are:
(1) Morrissey \etal (2007) -- GALEX,
(2) Bessell (1979) -- {\it UBVRI},
(3) Oke and Gunn 1983, Fukugita \etal (1996) -- SDSS,
(4) Cohen, Wheaton and Megeath (2003) -- 2MASS,
(5) Reach \etal (2005) -- IRAC,
(6) Rieke \etal (2008) -- MIPS 24~$\mu$m,
(7) Gordon \etal (2007) -- MIPS 70~$\mu$m,
(8) Stansberry \etal (2007) -- MIPS 160~$\mu$m, and
(9) Jarrett \etal (2011), Wright \etal (2010) -- WISE.}
}

While the NDWFS survey provides the {\it RI} magnitudes, it does not
provide the {\it V}-band data. We therefore calculate the $g-V$ synthetic
colors as a function of redshift to obtain the {\it V}-band magnitudes and
calibrate them with the observed $V-I$ colors based on 758 quasars from the
Magellanic Quasars Survey (Koz³owski \etal 2013), where the {\it V} and
{\it I}-band magnitudes were provided by the OGLE sky survey (Udalski \etal
2008, Udalski, Szymañski and Szymañski 2015).

The extinction-corrected monochromatic luminosity at 5100~\AA, 3000~\AA\ and\break
1350~\AA\ were already provided in Shen \etal (2011) for cosmological
parameters 
$H_0=70$~(km/s)/Mpc, $\Omega_{\rm M}=0.3$ and
$\Omega_\lambda=0.7$. We adopt the same cosmological parameters, hence our
broad-band--to--monochromatic luminosity ratios are cosmological
model-indepen\-dent and can be used with any other set of cosmological
parameters. One should be careful about using the low-$z$ ratios, as both
the broad-band filters and the monochromatic luminosity can be affected by
AGN host contamination.
\vspace*{-9pt}
\Section{Methodology}
\vspace*{-5pt}
A broad-band luminosity $L_F$ ($\nu L_\nu$) in a filter $F$ is calculated from
$$L_F=4\pi D_L^2\alpha_F\nu_F10^{-0.4\times m_F}\eqno(1)$$
where $L_F$ is in $10^{23}~$erg/s, $\alpha_F$ is the zero-magnitude flux
[Jy] for a given filter $F$ (column~3 in Table~1), $m_F$ is the observed
source magnitude, $\nu_F$ is the central frequency [Hz] for that filter
(column~4 in Table~1), and $D_L$ is the luminosity distance [cm]. Details
of the $D_L$ calculation\footnote{\it
  http://www.astro.ucla.edu/$\sim$wright/CosmoCalc.html} from a redshift
$z$ and cosmological parameters are given in Wright (2006).

Using the above prescription, we convert the extinction-corrected observed
broad-band {\it ugriz}, {\it VRI} (converted), {\it JHK} and WISE
magnitudes to the broad-band luminosities. We then calculate the empirical
ratios {\it R} of the broad-band--to--mono\-chromatic luminosity and provide
their medians along with dispersions and uncertainties in 0.01 redshift
bins (Table~2). They are also presented in Fig.~1. Note, that the
ratios hold independent of the cosmological model.
\begin{figure}[htb]
\hglue-1mm\includegraphics[width=6.6cm]{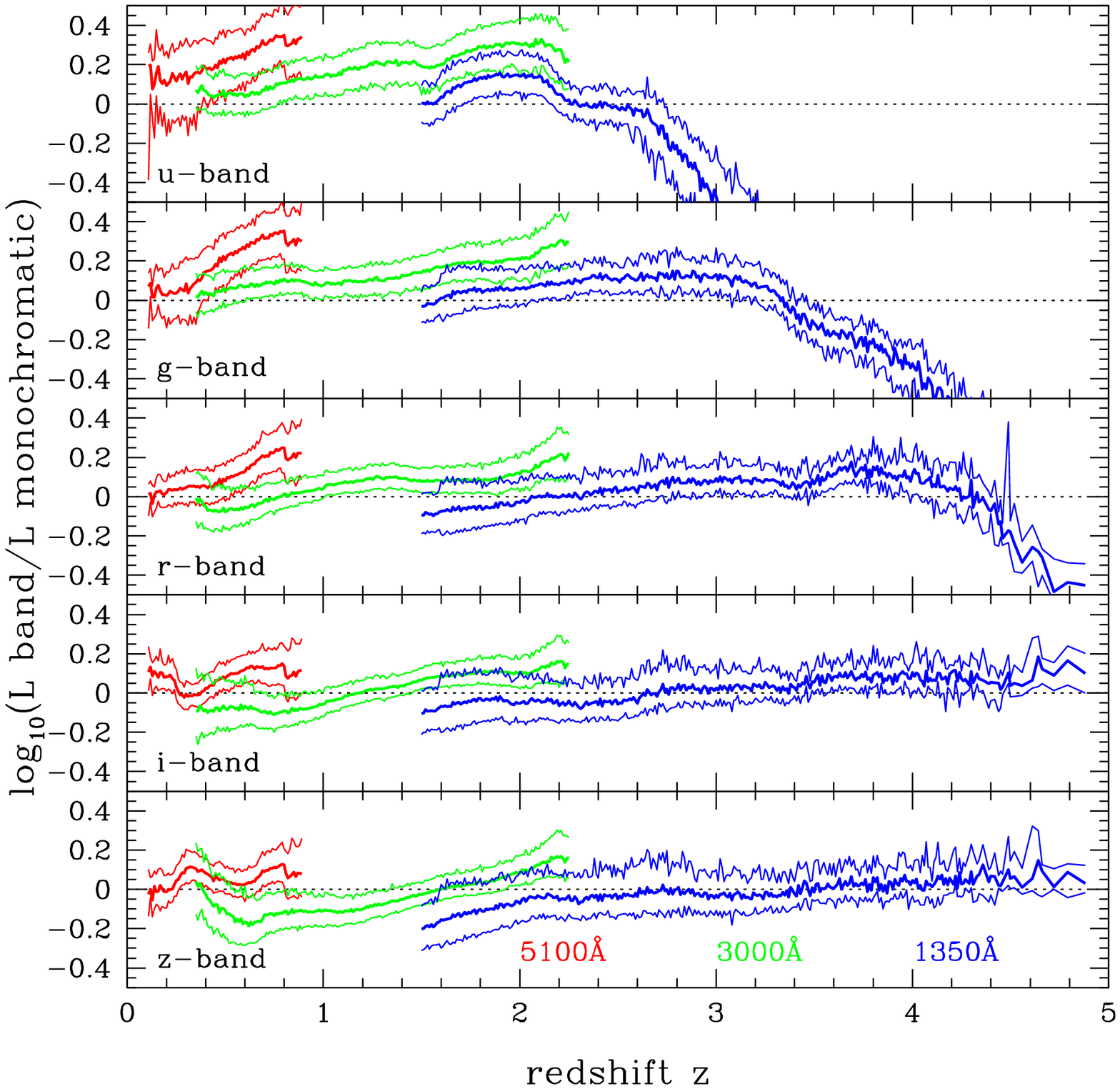}\includegraphics[width=6.5cm]{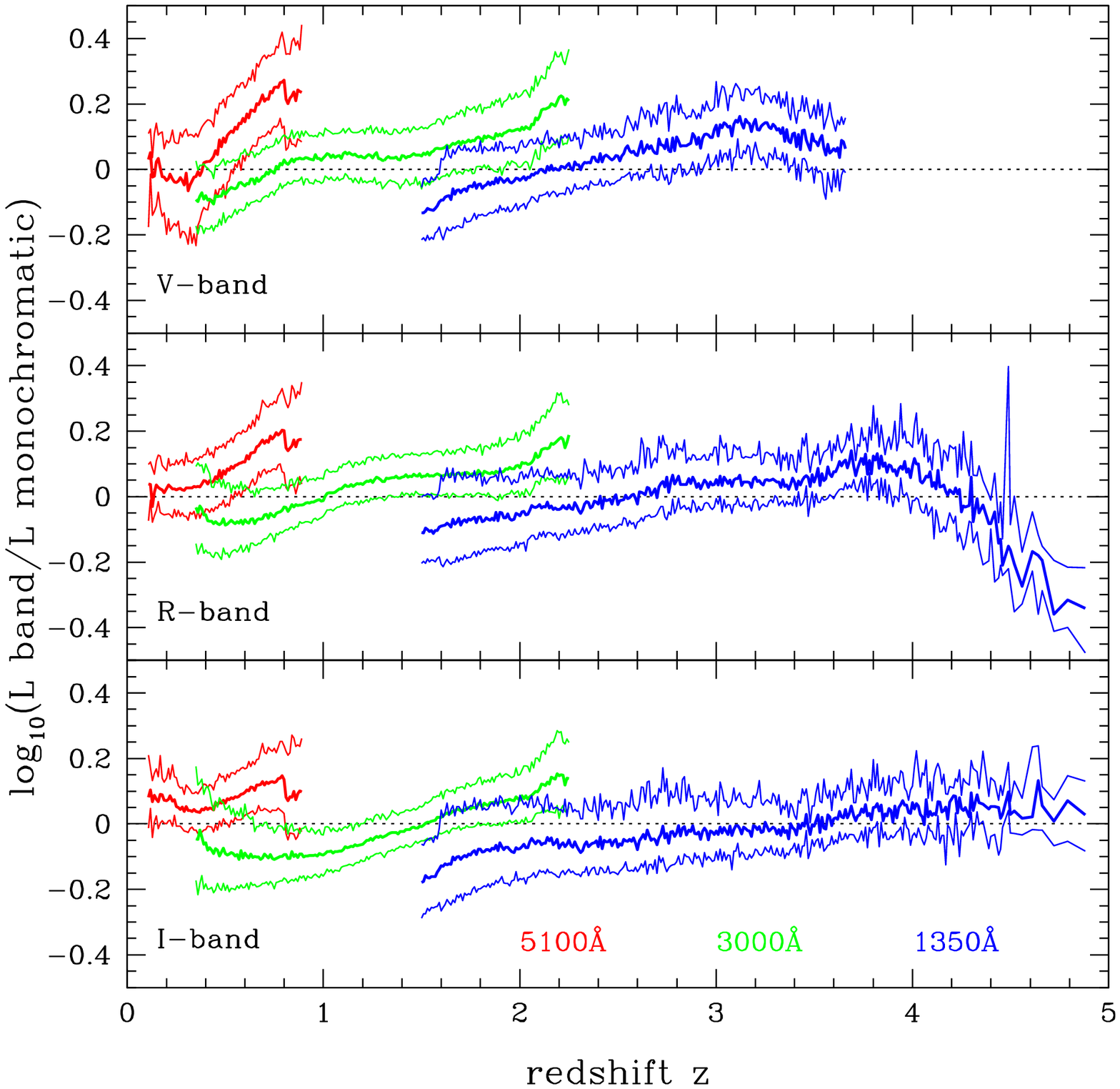}
\vskip9pt
\hglue-1mm\includegraphics[width=6.6cm]{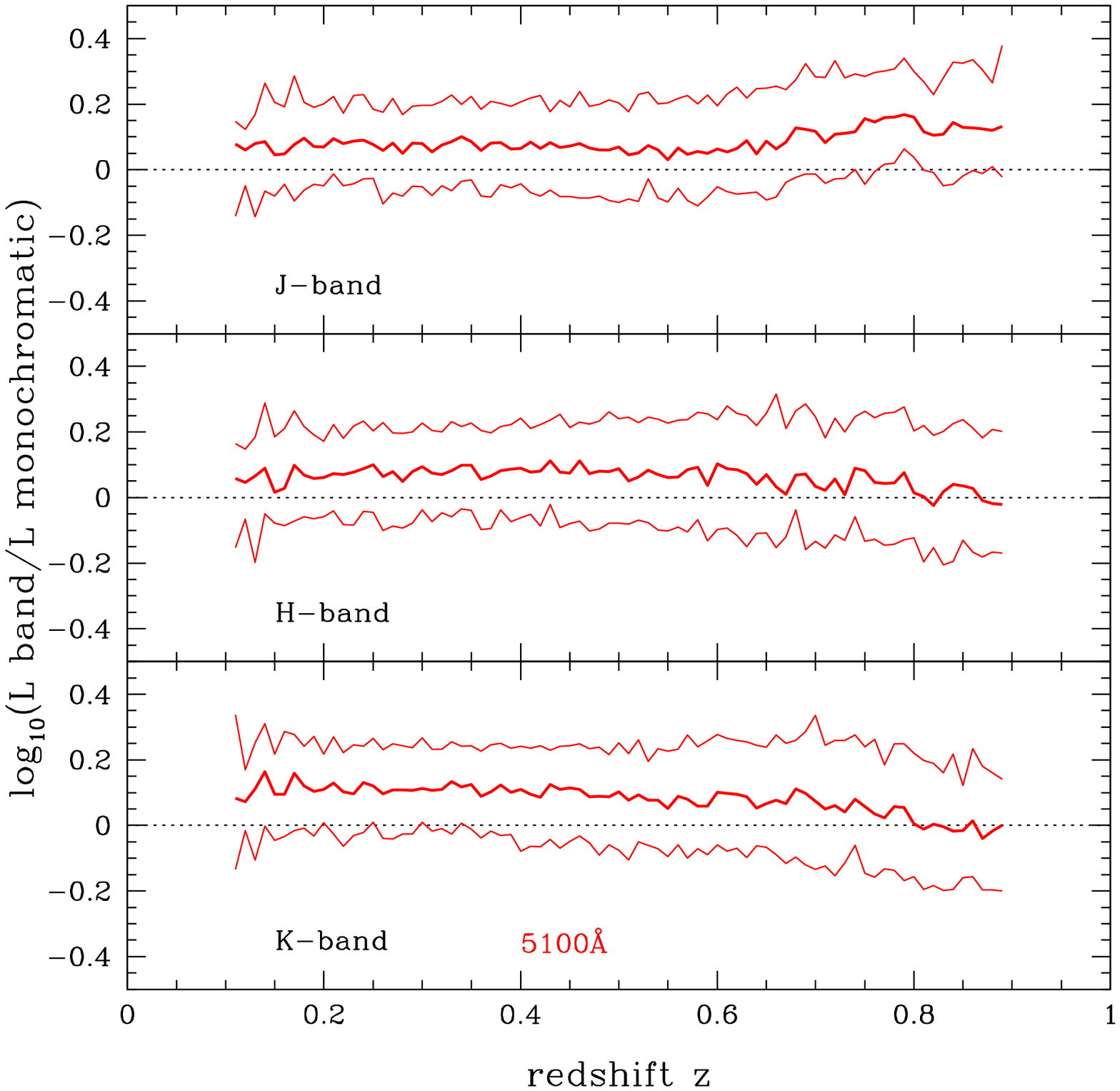}\includegraphics[width=6.5cm]{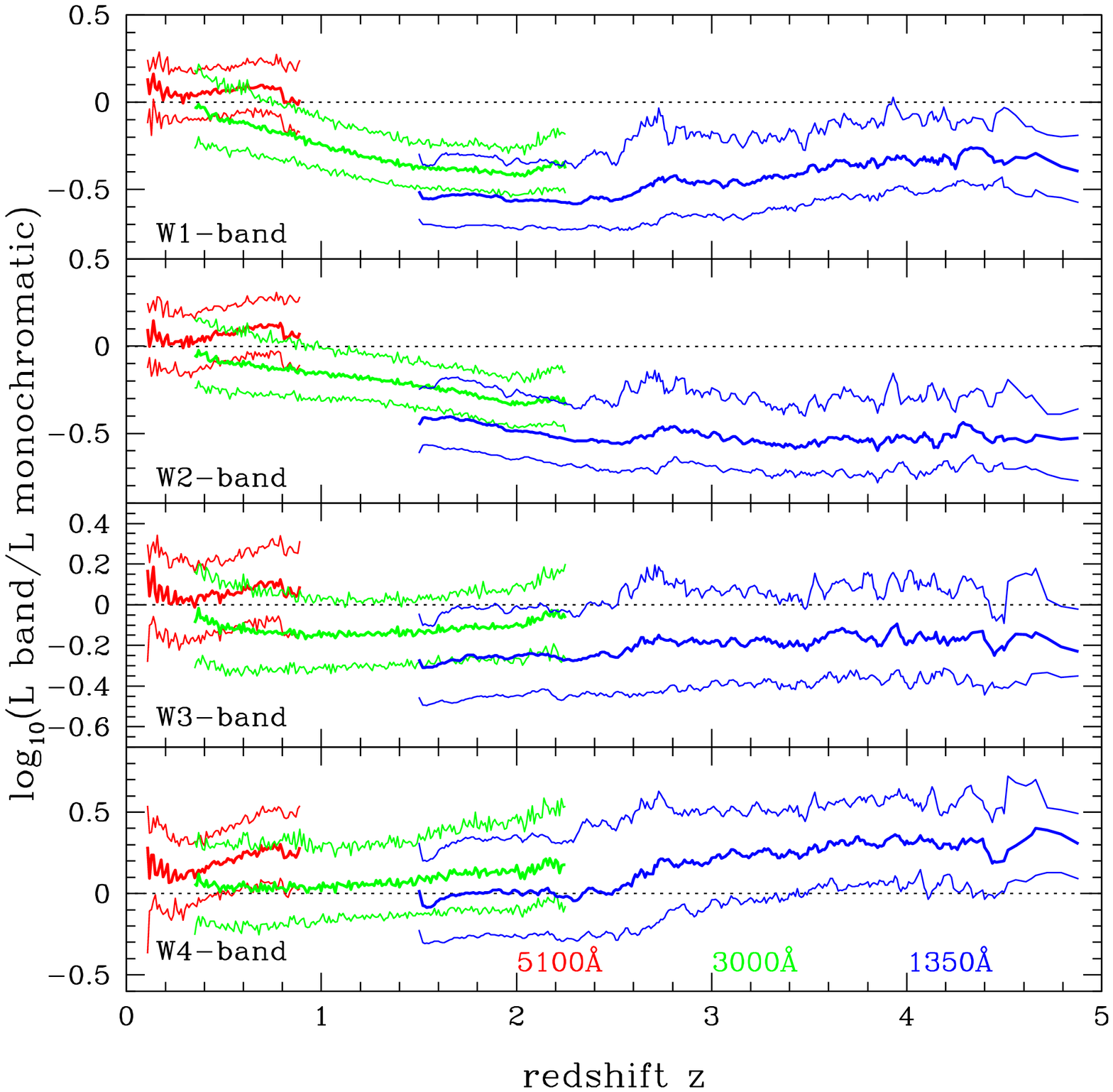}
\vskip7pt
\FigCap{Ratios between the broad-band SDSS {\it ugriz} ({\it top-left
    panel}), {\it VRI} ({\it top-right}), 2MASS {\it JHK (bottom-left)},
  and WISE ({\it bottom-right}) and monochromatic luminosity at 5100~\AA\
  (blue), 3000~\AA\ (green) and 1350~\AA\ (red). Both broad-band and
  monochromatic luminosity were derived with the same cosmological model,
  hence their ratios are model-independent. Having measured any of the
  magnitudes for an AGN at a redshift $z$, it is possible to convert that
  magnitude to either of three monochromatic luminosity with a typical
  dispersion of $\approx0.1$~dex. This may have profound implications for
  measurements of BLR radii and/or black hole masses.}
\end{figure}

\MakeTable{lc|rcc|cc|r}{12.5cm}{Median ratios between the broad-band and
monochromatic luminosity} 
{\hline
\noalign{\vskip3pt}
\multicolumn{8}{c}{at 5100~\AA}\\
\noalign{\vskip3pt}
\hline
\douprule
$F$ & $z$ & $\log_{10}(R)$ & $-\sigma$ & $\sigma$ &  ${\rm -err}$ & err & $N_{\rm obj}$\\
\hline
$u$ &  0.11 & 0.194 & $-0.582$ & 0.066 & $-0.141$ & 0.016 &  17 \\
$u$ &  0.12 & 0.199 & $-0.151$ & 0.106 & $-0.027$ & 0.019 &  32 \\
$u$ &  0.13 & 0.077 & $-0.149$ & 0.142 & $-0.028$ & 0.027 &  28 \\
$u$ &  0.14 & 0.098 & $-0.272$ & 0.131 & $-0.047$ & 0.023 &  33 \\
$u$ &  0.15 & 0.155 & $-0.113$ & 0.224 & $-0.017$ & 0.034 &  44 \\
\noalign{\vskip3pt}
\hline 
\noalign{\vskip3pt}
\multicolumn{8}{c}{at 3000~\AA}\\
\noalign{\vskip3pt}
\hline
\douprule
$F$ & $z$ & $\log_{10}(R)$ & $-\sigma$ & $\sigma$ &  ${\rm -err}$ & err & $N_{\rm obj}$\\
\hline
$u$ &  0.35 & 0.071 & $-0.085$ & 0.125 & $-0.006$ & 0.009 & 179 \\
$u$ &  0.36 & 0.056 & $-0.089$ & 0.114 & $-0.005$ & 0.006 & 336 \\
$u$ &  0.37 & 0.078 & $-0.101$ & 0.104 & $-0.006$ & 0.006 & 334 \\
$u$ &  0.38 & 0.094 & $-0.123$ & 0.103 & $-0.007$ & 0.006 & 290 \\
$u$ &  0.39 & 0.068 & $-0.096$ & 0.117 & $-0.005$ & 0.006 & 341 \\
\noalign{\vskip3pt}
\hline
\noalign{\vskip3pt}
\multicolumn{8}{c}{at 1350~\AA}\\
\noalign{\vskip3pt}
\hline
\douprule
$F$ & $z$ & $\log_{10}(R)$ & $-\sigma$ & $\sigma$ &  ${\rm -err}$ & err & $N_{\rm obj}$\\
\hline
\noalign{\vskip3pt}
$u$ &  1.50 &    0.001 & $-0.094$ & 0.091 & $-0.005$ & 0.005 & 353 \\
$u$ &  1.51 &    0.003 & $-0.103$ & 0.114 & $-0.004$ & 0.004 & 644 \\
$u$ &  1.52 & $-0.002$ & $-0.099$ & 0.120 & $-0.004$ & 0.005 & 574 \\
$u$ &  1.53 &    0.011 & $-0.109$ & 0.097 & $-0.005$ & 0.004 & 511 \\
$u$ &  1.54 &    0.001 & $-0.114$ & 0.086 & $-0.004$ & 0.003 & 717 \\
\noalign{\vskip3pt}
\hline
\noalign{\vskip3pt}
\multicolumn{8}{p{8.5cm}}{$F$ is the broad-band filter, $z$ is the redshift,
$\log_{10}(R)$ is the base 10 logarithm of the ratio between the band and
monochromatic continuum luminosity, $\sigma$ is the dispersion around the
median value, and ``err'' is the uncertainty of the median ratio. 
\newline
Full Table~2 is available in the electronic form from {\it Acta Astronomica
Archive}. A portion is shown here for guidance regarding its form
and content.}}

In addition to the tabular form, we also fit $\log_{10}(R)$ as a simple
function of redshift $z$ with the formula
$$\log_{10}(R)=a+bz\eqno(2)$$
in redshift ranges where this dependence is nearly linear. Typical
dispersions between the best fits and the tabular data are $\approx0.02$~dex
(\ie lower than the formal uncertainties from the tabular form). The
best-fit values for selected redshift ranges are provided in Tables~3--4
and the residuals between the measured median ratios and the fitted ones
are presented in Fig.~2.

\begin{figure}[htb]
\includegraphics[width=6.3cm]{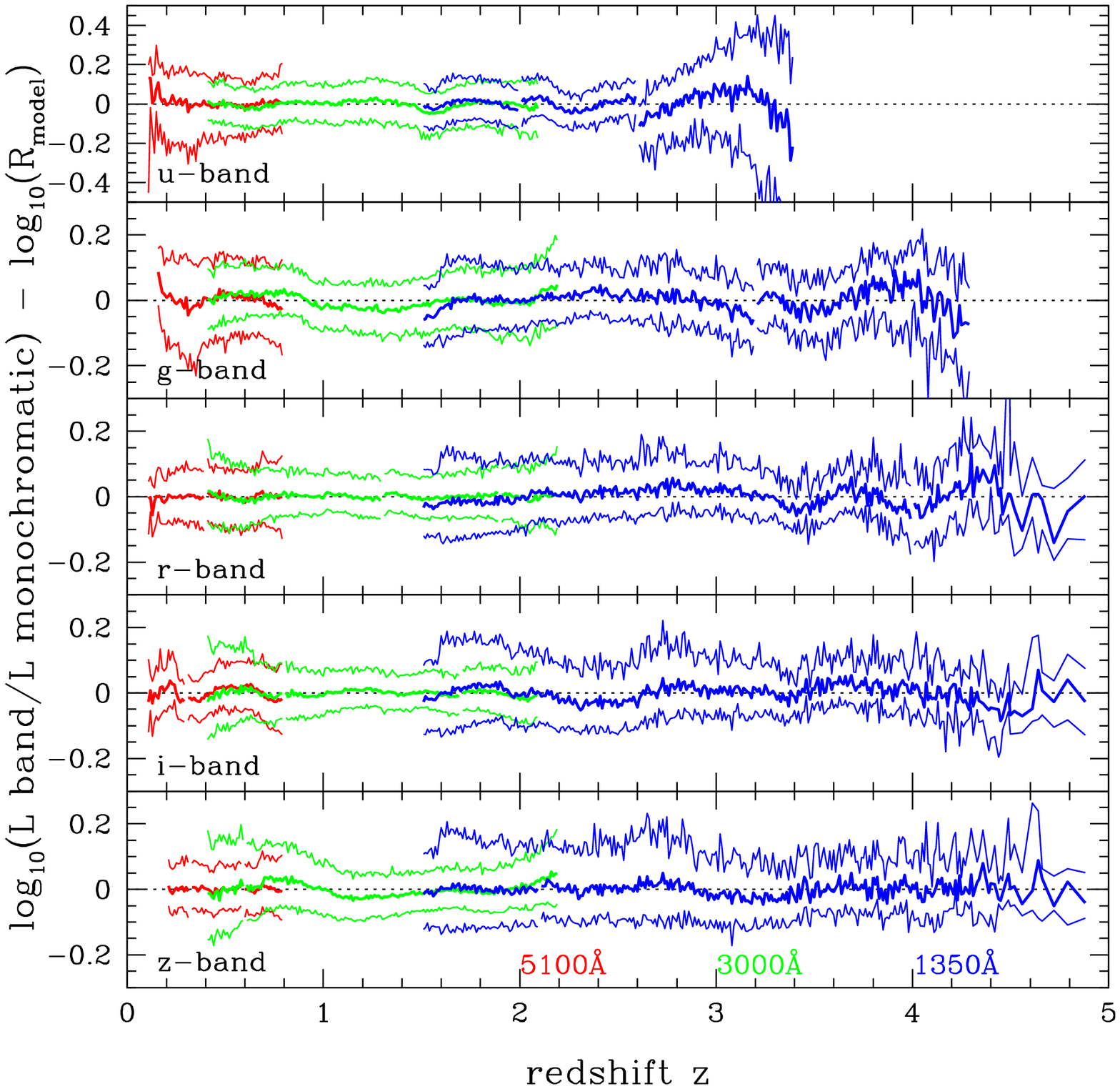}\includegraphics[width=6.3cm]{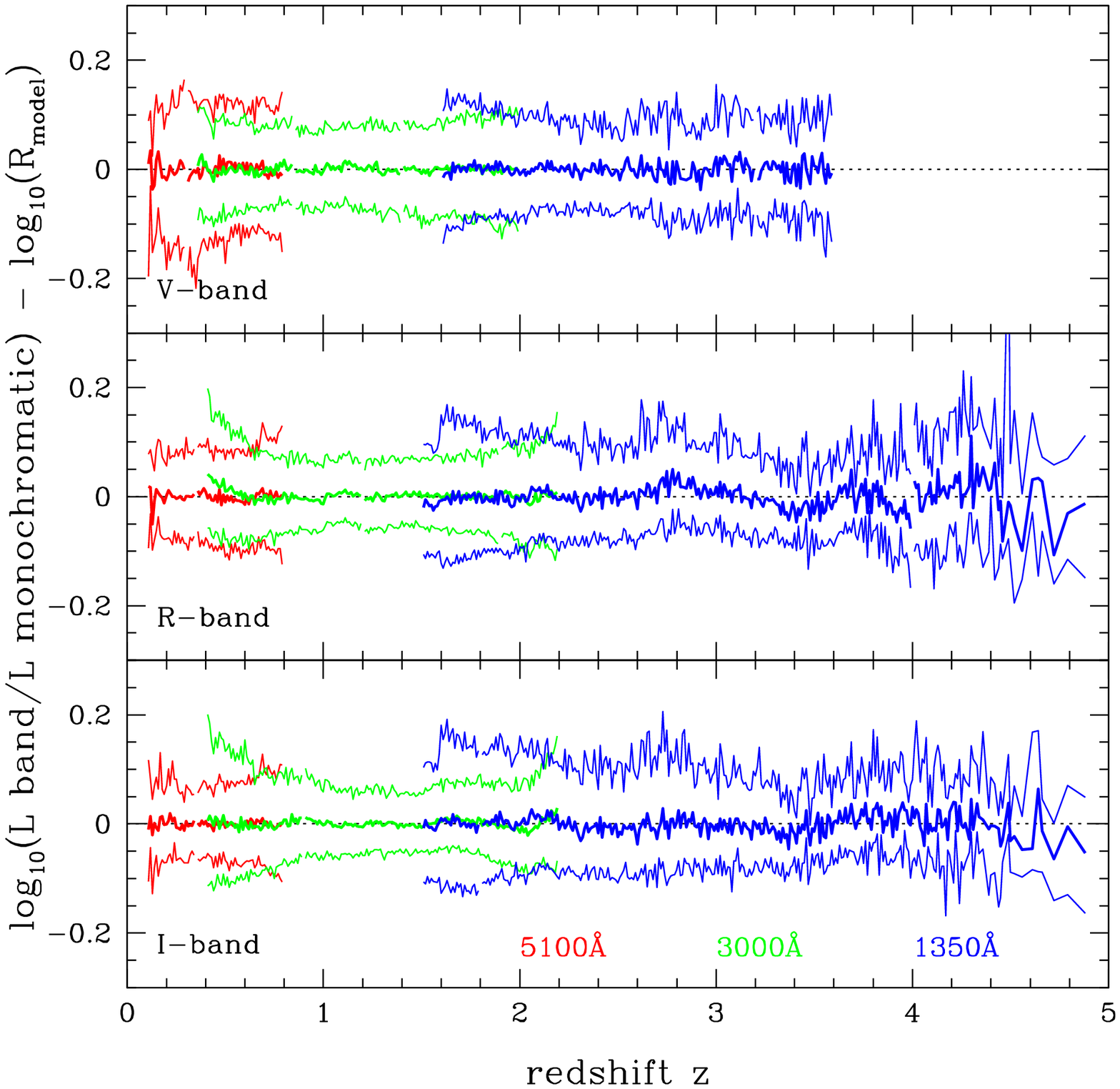}\\
\includegraphics[width=6.3cm]{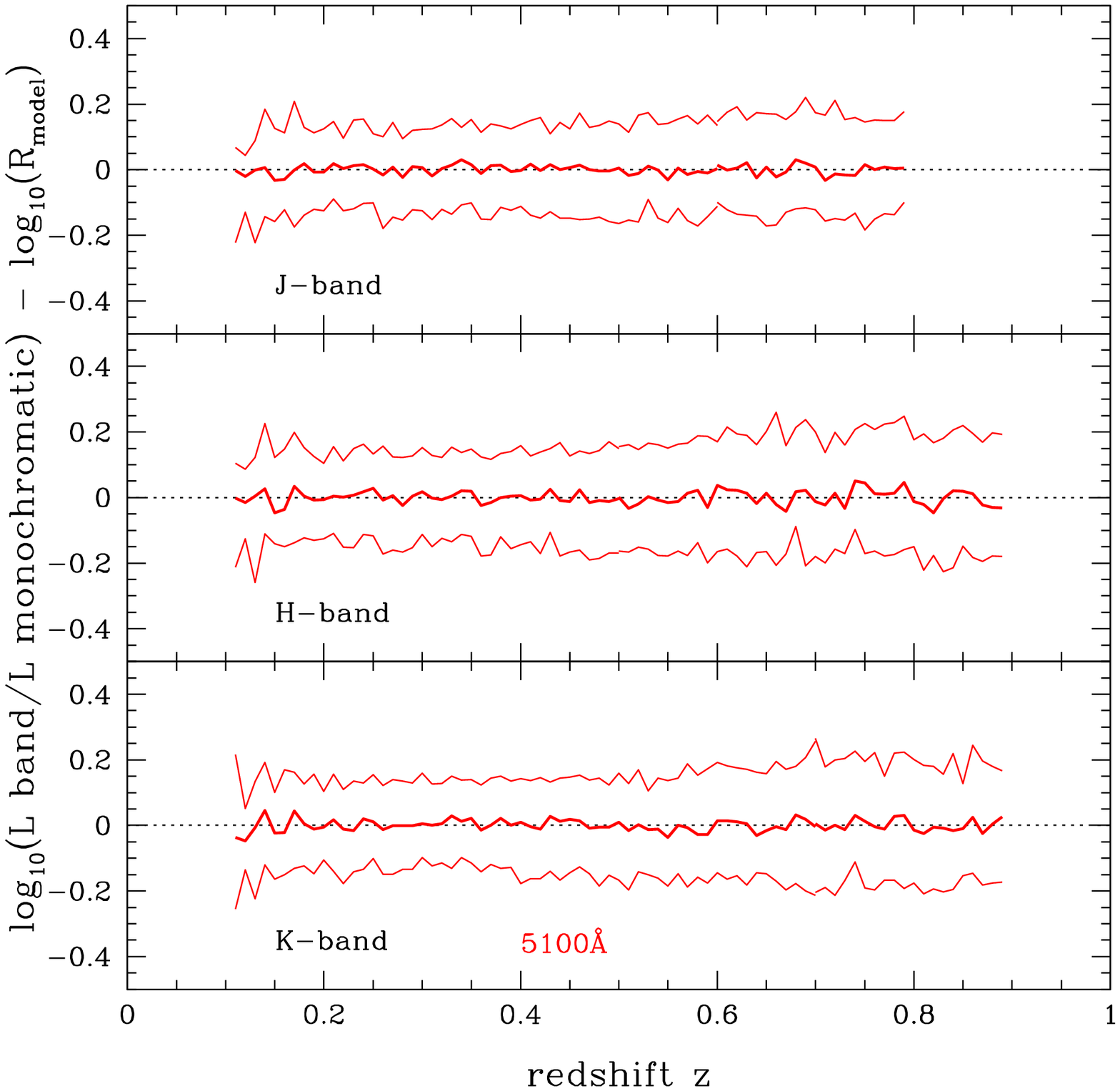}\includegraphics[width=6.3cm]{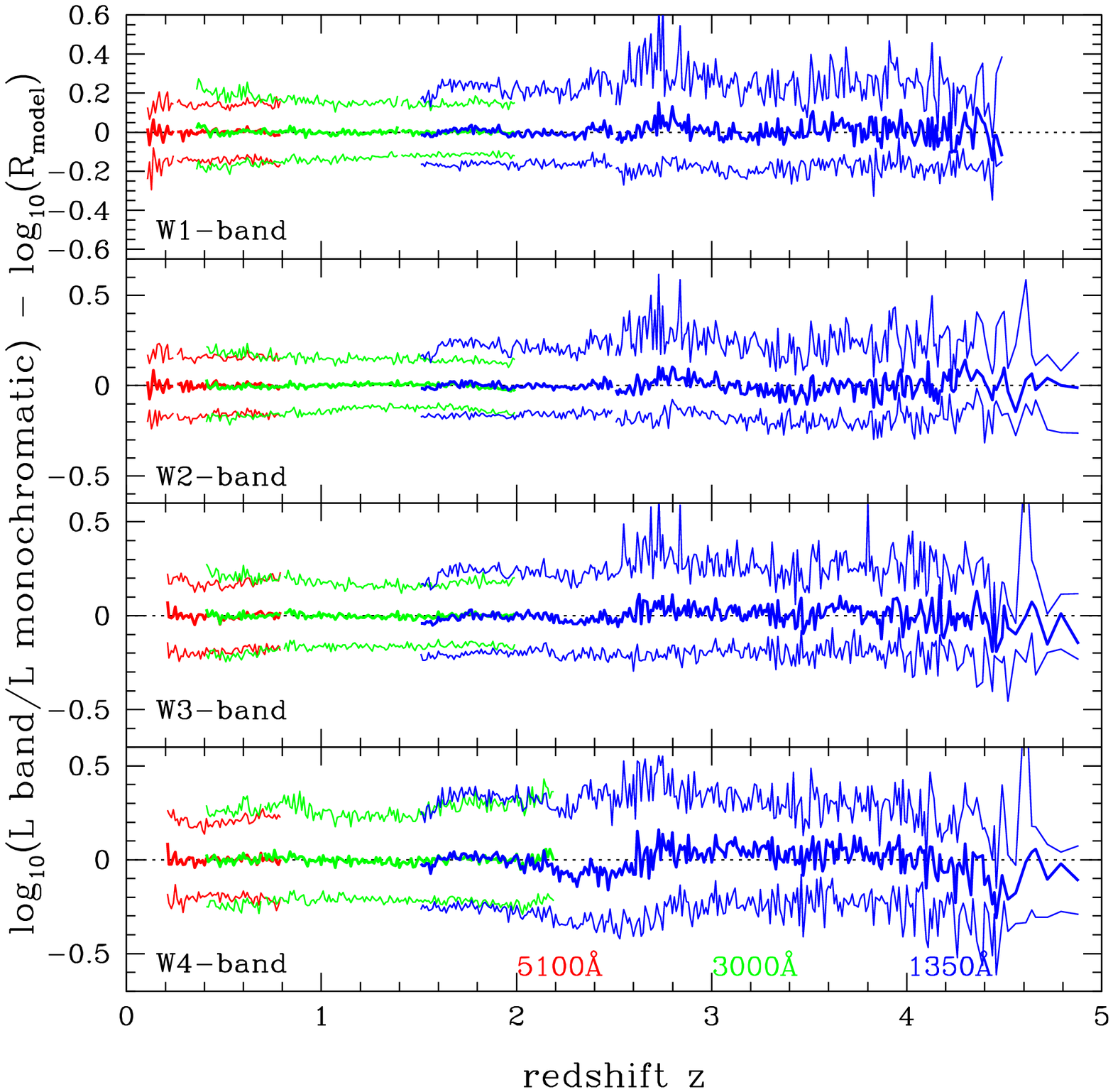}
\vskip5pt
\FigCap{Residuals after taking out the linear trends (Eq.~2) from the
  measured median ratios for the broad-band SDSS {\it ugriz} ({\it top-left
    panel}), {\it VRI} ({\it top-right}), 2MASS {\it JHK} ({\it
    bottom-left}), and WISE ({\it bottom-right}) and monochromatic
  luminosity at 5100~\AA\ (blue), 3000~\AA\ (green) and 1350~\AA\ (red).}
\end{figure}

A desired monochromatic luminosity can be estimated from
$$\log_{10}(\lambda L_\lambda)=\log_{10}(L_F)-\log_{10}(R)\eqno(3)$$
where $\log_{10}(L_F)$ is estimated from Eq.(1) and $\log_{10}(R)$ is
provided either in Table~2 or parametric form from Eq.(2) with the
fitted values stored in Tables~3--4.

\renewcommand{\TableFont}{\scriptsize}
\MakeTableSepp{lc|c|rrrr}{12.5cm}{Best-fit parameters to Eq.(2) for optical bands}
{\hline
\douprule
$F$ & $\lambda_{\rm cont.}$ & $z$ range & $a$ & $\sigma_a$ & $b$ &$\sigma_b$ \\
\hline
\uprule
$u$ &  5100~\AA &  0.1--0.8 &	0.019  & 0.006 &    0.405 & 0.011 \\
$g$ &  5100~\AA & 0.15--0.8 & $-$0.107 & 0.006 &    0.613 & 0.013 \\
$r$ &  5100~\AA &  0.1--0.4 &    0.002 & 0.005 &    0.142 & 0.017 \\
$r$ &  5100~\AA &  0.4--0.8 & $-$0.194 & 0.008 &    0.562 & 0.012 \\
$i$ &  5100~\AA &  0.1--0.3 &	0.212  & 0.019 & $-$0.719 & 0.093 \\
$i$ &  5100~\AA &  0.3--0.8 & $-$0.111 & 0.010 &    0.352 & 0.018 \\
$z$ &  5100~\AA & 0.2--0.32 & $-$0.271 & 0.019 &    1.245 & 0.074 \\ 
$z$ &  5100~\AA & 0.32--0.59&    0.231 & 0.009 & $-$0.373 & 0.019 \\ 
$z$ &  5100~\AA & 0.59--0.8 & $-$0.323 & 0.020 &    0.578 & 0.029 \\ 
\hline
\uprule
$u$ &  3000~\AA &  0.4--2.1 & $-$0.037 & 0.003 &    0.175 & 0.002 \\
$g$ &  3000~\AA &  0.4--2.2 & $-$0.020 & 0.004 &    0.120 & 0.003 \\
$r$ &  3000~\AA &  0.4--1.3 & $-$0.178 & 0.003 &    0.222 & 0.003 \\
$r$ &  3000~\AA &  1.3--1.9 &    0.106 & 0.008 & $-$0.013 & 0.005 \\
$r$ &  3000~\AA &  1.9--2.2 & $-$0.654 & 0.036 &    0.389 & 0.017 \\
$i$ &  3000~\AA &  0.4--0.8 & $-$0.051 & 0.010 & $-$0.057 & 0.016 \\
$i$ &  3000~\AA &  0.8--1.7 & $-$0.279 & 0.004 &    0.229 & 0.003 \\
$i$ &  3000~\AA &  1.7--2.2 & $-$0.050 & 0.024 &    0.083 & 0.012 \\
$z$ &  3000~\AA &  0.4--0.6 &	0.316  & 0.021 & $-$0.852 & 0.045 \\ 
$z$ &  3000~\AA &  0.6--2.2 & $-$0.299 & 0.005 &    0.190 & 0.003 \\ 
\hline
\uprule
$u$ &  1350~\AA &  1.5--2.0 & $-$0.513 & 0.032 &    0.348 & 0.018 \\
$u$ &  1350~\AA &  2.0--2.6 &	0.810  & 0.047 & $-$0.337 & 0.020 \\
$u$ &  1350~\AA &  2.6--3.4 &	4.295  & 0.113 & $-$1.616 & 0.038 \\
$g$ &  1350~\AA &  1.5--3.2 & $-$0.074 & 0.009 &    0.068 & 0.004 \\
$g$ &  1350~\AA &  3.2--4.3 &	1.961  & 0.049 & $-$0.587 & 0.013 \\
$r$ &  1350~\AA &  1.5--4.0 & $-$0.188 & 0.006 &    0.083 & 0.002 \\
$r$ &  1350~\AA &  4.0--4.9 &	2.985  & 0.156 & $-$0.705 & 0.036 \\
$i$ &  1350~\AA &  1.5--4.9 & $-$0.165 & 0.005 &    0.060 & 0.002 \\
$z$ &  1350~\AA &  1.5--2.1 & $-$0.576 & 0.012 &    0.259 & 0.007 \\ 
$z$ &  1350~\AA &  2.1--4.9 & $-$0.149 & 0.007 &    0.045 & 0.002 \\ 
\hline
\uprule
$V$ &  5100~\AA & 0.1--0.3  &	0.062  & 0.017 & $-$0.381 & 0.081 \\
$V$ &  5100~\AA & 0.3--0.8  & $-$0.249 & 0.006 &    0.667 & 0.011 \\
$R$ &  5100~\AA & 0.1--0.35 &	0.017  & 0.009 &    0.030 & 0.040 \\
$R$ &  5100~\AA & 0.35--0.8 & $-$0.128 & 0.006 &    0.415 & 0.010 \\
$I$ &  5100~\AA & 0.1--0.35 &	0.118  & 0.007 & $-$0.234 & 0.030 \\
$I$ &  5100~\AA & 0.35--0.8 & $-$0.057 & 0.004 &    0.255 & 0.006 \\ 
\hline
\uprule
$V$ &  3000~\AA & 0.35--0.85& $-$0.199 & 0.006 &    0.263 & 0.009 \\
$V$ &  3000~\AA & 0.85--1.4 &    0.035 & 0.006 &    0.002 & 0.006 \\
$V$ &  3000~\AA & 1.4--2.0  & $-$0.175 & 0.006 &    0.149 & 0.003 \\
$R$ &  3000~\AA & 0.4--1.2  & $-$0.195 & 0.004 &    0.196 & 0.004 \\
$R$ &  3000~\AA & 1.2--1.9  &    0.007 & 0.004 &    0.036 & 0.003 \\
$R$ &  3000~\AA & 1.9--2.2  & $-$0.580 & 0.036 &    0.339 & 0.017 \\
$I$ &  3000~\AA & 0.4--0.9  & $-$0.072 & 0.005 & $-$0.036 & 0.008 \\
$I$ &  3000~\AA & 0.9--2.2  & $-$0.275 & 0.003 &    0.182 & 0.018 \\
\hline
\uprule
$V$ &  1350~\AA & 1.6--3.2  & $-$0.301 & 0.005 &    0.138 & 0.002 \\
$V$ &  1350~\AA & 3.2--3.6  &	0.778 & 0.083 & $-$0.199 & 0.024 \\
$R$ &  1350~\AA & 1.5--4.0  & $-$0.218 & 0.004 &    0.084 & 0.002 \\
$R$ &  1350~\AA & 4.0--4.9  &    2.009 & 0.092 & $-$0.479 & 0.022 \\
$I$ &  1350~\AA & 1.5--1.8  & $-$0.671 & 0.028 &    0.333 & 0.017 \\
$I$ &  1350~\AA & 1.8--4.9  & $-$0.182 & 0.005 &    0.054 & 0.002 \\
\hline}

\renewcommand{\TableFont}{\footnotesize}
\MakeTableee{lc|c|rrrr}{12.5cm}{Best-fit parameters to Eq.(2) for infrared bands}
{\hline
\douprule
$F$ & $\lambda_{\rm cont.}$ & $z$ range & $a$ & $\sigma_a$ & $b$ &$\sigma_b$ \\
\hline
\uprule
$J$ &  5100~\AA &  0.1--0.6 &	0.085 & 0.005 & $-$0.042 & 0.014 \\
$J$ &  5100~\AA &  0.6--0.8 & $-$0.313& 0.051 &    0.603 & 0.073 \\
$H$ &  5100~\AA &  0.1--0.5 &	0.051 & 0.008 &    0.081 & 0.025 \\
$H$ &  5100~\AA &  0.5--0.9 &	0.184 & 0.026 & $-$0.196 & 0.037 \\
$K$ &  5100~\AA &  0.1--0.7 &	0.128 & 0.006 & $-$0.071 & 0.015 \\
$K$ &  5100~\AA &  0.7--0.9 &	0.422 & 0.066 & $-$0.503 & 0.082 \\ 
\hline
\uprule
$W1$&  5100~\AA & 0.1--0.25 &	0.191 & 0.042 & $-$0.655 & 0.232 \\
$W1$&  5100~\AA & 0.25--0.8 &	0.007 & 0.006 &    0.112 & 0.010 \\
$W2$&  5100~\AA & 0.1--0.25 &	0.127 & 0.046 & $-$0.460 & 0.258 \\
$W2$&  5100~\AA & 0.25--0.8 & $-$0.040& 0.006 &    0.217 & 0.012 \\
$W3$&  5100~\AA & 0.2--0.8  & $-$0.022& 0.008 &    0.162 & 0.015 \\
$W4$&  5100~\AA & 0.2--0.8  &	0.013 & 0.008 &    0.361 & 0.016 \\ 
\hline
\uprule
$W1$&  3000~\AA & 0.35--1.4 &	0.054  & 0.003& $-$0.295 & 0.004 \\
$W1$&  3000~\AA & 1.4--2.0  & $-$0.242 & 0.010& $-$0.085 & 0.006 \\
$W2$&  3000~\AA & 0.4--2.0  & $-$0.005 & 0.002& $-$0.152 & 0.002 \\
$W3$&  3000~\AA & 0.4--0.8  & $-$0.030 & 0.012& $-$0.146 & 0.020 \\
$W3$&  3000~\AA & 0.8--2.0  & $-$0.186 & 0.004&    0.039 & 0.003 \\
$W4$&  3000~\AA & 0.4--0.7  &	0.080  & 0.022& $-$0.071 & 0.040 \\
$W4$&  3000~\AA & 0.7--2.2  & $-$0.058 & 0.006&    0.100 & 0.003 \\ 
\hline
\uprule
$W1$&  1350~\AA & 1.5--2.5  & $-$0.488 & 0.013& $-$0.033 & 0.006 \\
$W1$&  1350~\AA & 2.5--4.5  & $-$0.846 & 0.026&    0.127 & 0.007 \\
$W2$&  1350~\AA & 1.5--2.5  & $-$0.131 & 0.011& $-$0.174 & 0.005 \\
$W2$&  1350~\AA & 2.5--4.9  & $-$0.511 & 0.021& $-$0.005 & 0.006 \\
$W3$&  1350~\AA & 1.5--4.9  & $-$0.338 & 0.010&    0.044 & 0.003 \\
$W4$&  1350~\AA & 1.5--4.9  & $-$0.274 & 0.014&    0.146 & 0.005 \\
\hline
\noalign{\vskip5pt}
\multicolumn{7}{p{7.9cm}}{2MASS detections are limited to $z<0.8$ sources
therefore the measurement of conversions to monochromatic luminosities at
3000~\AA\ and 1350~\AA\ is not feasible.}}

\Section{A Simple Prescription}
Imagine a situation when one knows that the source is an AGN but its
spectrum is not flux-calibrated or its signal-to-noise (S/N) is too
low to reliably measure its monochromatic luminosity. What one has at
hand are just its broad-band common magnitudes and the redshift
estimate.

As an example, let assume we are interested in obtaining the monochromatic
luminosity for a quasar SDSS J000006.53+003055.2 at (RA,
Decl.)~=~(0.027228, 0.551534)~deg at a redshift of $z=1.8246$. It is drawn
from the sample analy\-zed in the paper, hence the true spectral
monochromatic luminosity is known. Its extinction-corrected magnitudes are
$u=20.254\pm0.065$~mag, $g=20.365\pm0.034$~mag, $r=20.255\pm0.038$~mag,
$i=20.040\pm0.041$~mag, $z=20.005\pm0.121$~mag, $V=20.431$~mag (estimated),
$R=20.067$~mag (estimated), $I=19.661$ mag (estimated), {\it JHK} not
measured, $W1=16.560\pm0.084$~mag, $W2=15.094\pm0.093$~mag,
$W3=12.549\pm0.539$~mag, and $W4=8.072$~mag.

The broad-band luminosities [erg/s], are then
$\log_{10}{(L_u)}=45.748\pm0.026$, $\log_{10}{(L_g)}=45.583\pm0.014$,
$\log_{10}{(L_r)}=45.508\pm0.015$, $\log_{10}{(L_i)}=45.510\pm0.016$,
$\log_{10}{(L_z)}=45.447\pm0.048$, $\log_{10}{(L_V)}=45.485$,
$\log_{10}{(L_R)}=45.495$, $\log_{10}{(L_I)}=45.462$,
$\log_{10}{(L_{W1})}=45.177\pm0.034$, $\log_{10}{(L_{W2})}=45.370\pm0.037$,
$\log_{10}{(L_{W3})}=45.271\pm0.216$, and $\log_{10}{(L_{W4})}=46.185$.

From Fig.~1, it is clear that at this redshift it is possible to derive
both 3000~\AA\ and 1350~\AA\ luminosity from the broad-band filters.
Table~5 gives the derived values. In this particular case, the optical
bands seem to reproduce the real values better (to within $\approx1\sigma$)
than the infrared ones. The overestimate of the luminosity from the
infrared filters may be attributed to the host contamination, as galaxies
are usually much brighter in infrared than in visible light.

\renewcommand{\TableFont}{\footnotesize}
\MakeTable{l|rc|rc}{12.5cm}{Example of monochromatic luminosity estimates 
for AGN SDSS J000006.53+003055.2}
{\hline
\uprule
filter & $\log_{10}(R)$ & $\log_{10}(\lambda L_\lambda/{\rm erg/s})$& $\log_{10}(R)$ & $\log_{10}(\lambda L_\lambda/{\rm erg/s}$\\
 & & at 3000~\AA & & at 1350~\AA \\
\hline
\douprule
spec. & $\cdots$ & $45.32\pm0.04$ & $\cdots$ & $45.60\pm0.03$ \\
\hline
\uprule
$u$  & $ 0.30$ & $45.45\pm0.12$ & $ 0.14$ & $45.61\pm0.11$ \\
$g$  & $ 0.20$ & $45.38\pm0.09$ & $ 0.05$ & $45.53\pm0.10$ \\
$r$  & $ 0.09$ & $45.42\pm0.07$ & $-0.06$ & $45.57\pm0.11$ \\
$i$  & $ 0.11$ & $45.40\pm0.07$ & $-0.04$ & $45.55\pm0.12$ \\
$z$  & $ 0.04$ & $45.41\pm0.07$ & $-0.11$ & $45.56\pm0.13$ \\ \hline\uprule
$V$  & $ 0.10$ & $45.39\pm0.09$ & $-0.05$ & $45.54\pm0.10$ \\ 
$R$  & $ 0.08$ & $45.42\pm0.07$ & $-0.07$ & $45.57\pm0.11$ \\
$I$  & $ 0.06$ & $45.40\pm0.07$ & $-0.08$ & $45.54\pm0.12$ \\ \hline\uprule
$W1$ & $-0.40$ & $45.58\pm0.13$ & $-0.54$ & $45.72\pm0.20$ \\ 
$W2$ & $-0.29$ & $45.66\pm0.14$ & $-0.45$ & $45.82\pm0.19$ \\
$W3$ & $-0.11$ & $45.38\pm0.16$ & $-0.28$ & $45.55\pm0.20$ \\
$W4$ & $ 0.14$ & $46.05\pm0.28$ & $ 0.00$ & $46.19\pm0.30$ \\
\hline}

\Section{Discussion}
AGN are well known as variable sources (see Ulrich, Maraschi and Urry 1997
for a review). Their variability is aperiodic and well-modeled by the
damped random walk method (\eg Kelly, Bechtold and Siemiginowska 2009,
Koz³owski \etal 2010a, Zu \etal 2013a). A simplified description of their
variability is {\it via} the structure function -- a quantity that measures
the average magnitude difference for a time difference between any two
epochs (\eg Schmidt \etal 2010). It is described by
$$SF(\tau)=SF_0\left(\frac{\tau}{\tau_0}\right)^\gamma\eqno(4)$$
where $SF_0$ is the structure function at a fixed $\tau_0$, $\tau$ is the
time difference between two observations, and $\gamma$ is the slope of the
structure function.

Because both the broad-band optical or infrared observations and the
spectra were taken at different epochs, usually a few years apart, a
fraction of the scatter in the derived ratios is due to variability itself
and not the intrinsic differences between AGNs. Collecting information on
the time differences between all observations is beyond the scope of this
paper, nevertheless we roughly estimate the imprint of variability
contribution to the scatter in the relations.

Most SDSS observations investigated here, both spectroscopic and
photometric, were obtained in the years 2000--2010, as were the WISE
observations. Only the 2MASS data were obtained earlier. For an order of
magnitude estimate, we assume that the median difference between any two
observations is five years, and we know from the sample that the median AGN
redshift is $z\approx1.5$. This means that the median rest-frame time
difference is five years times $(1+z)^{-1}$, hence two years.

Vanden Berk \etal (2004) measured the {\it gri} structure function
parameters for 25\,000 SDSS quasars with approximately $\tau_0=2$~years,
$SF_0=0.28$~mag and $\gamma=0.30$. Since our median rest-frame time
difference is two years and both the broad-band and monochromatic light
curves are highly correlated, we can expect the median change in
$\log_{10}(R)$ to be of order of 0.1~dex. This implies that this
variability is responsible for a significant fraction of the dispersions
reported in this paper.

Koz³owski \etal (2010b, 2015) analyzed the variability of $\approx1500$
quasars observed by the Spitzer Space Telescope at 3.6~$\mu$m and
4.5~$\mu$m (bands nearly identical to the $W1$ and $W2$ WISE bands,
respectively), and found for the rest-frame $\tau_0=2$~years
$SF_0=0.12$~mag and $\gamma=0.47$ for 3.6~$\mu$m and $SF_0=0.13$~mag and
$\gamma=0.45$ for 4.5~$\mu$m. Again, we estimate the expected median
variance between the optical and IR bands in $\log_{10}(R)$ is $<0.1$~dex.

It is clear that such intrinsic AGN parameters as the black hole mass,
emission lines strength and width, spectrum shape, or the host
contamination will have impact on the derived ratios. In Fig.~3, we inspect
the dependence of the derived ratios on the emission line width (top-left
panel), its strength (top-right panel), the black hole mass (bottom-left
panel), and the spectral slope (bottom-right panel) using the estimates
from Shen \etal (2011). As an example, we use the conversion from the
$r$-band to monochromatic luminosity at 3000~\AA\ and the MgII line
parameters for the $z=0.4{-}2.1$ quasars. From Fig.~3, we see that there is
little or no dependence of $\log_{10}(R)$ as the MgII line width and
strength or the black hole mass. There is a hint of weak anti-correlations
with $\log_{10}(R)$ for the latter two observables, but they are most
prominent at the redshift extremes, where there are less objects per bin,
and the ratios are less precisely determined. There is, as expected,
correlation with the continuum slope. Over 91\% of AGNs have slopes in the
range $-2<{\rm slope}<0$, while the ratio changes by up to $1\sigma$ inside
the entire redshift range between $-2<{\rm slope}<-1$ and $-1<{\rm
slope}<0$. The ratios for AGNs with $0<{\rm slope}<2$ seem not to follow
the ones from the main sample, they, however, constitute only 2\% of the
sample.

The BLR radius-relation is $r\propto L^{1/2}$, so the black hole mass
scales as $M_{BH}\propto L^{1/2} v^2$. Keeping the BLR velocity fixed, an
0.1~dex change in luminosity introduces $\approx12\%$ change in radius
and/or black hole mass. These unlucky 2\% of AGNs with $0<{\rm slope}<2$
will introduce additional biases of that order in the black hole mass or
BLR radius estimate.

We note that these luminosity conversions are based on a large sample of
``average'' quasars. There are some residual correlations but they appear
to introduce typical shift of 0.1~dex -- a minor price to pay if the
spectrum for the object does not exist, is of low quality, or it is not
feasible to determine the continuum slope.

\begin{figure}[htb]
\includegraphics[width=12cm]{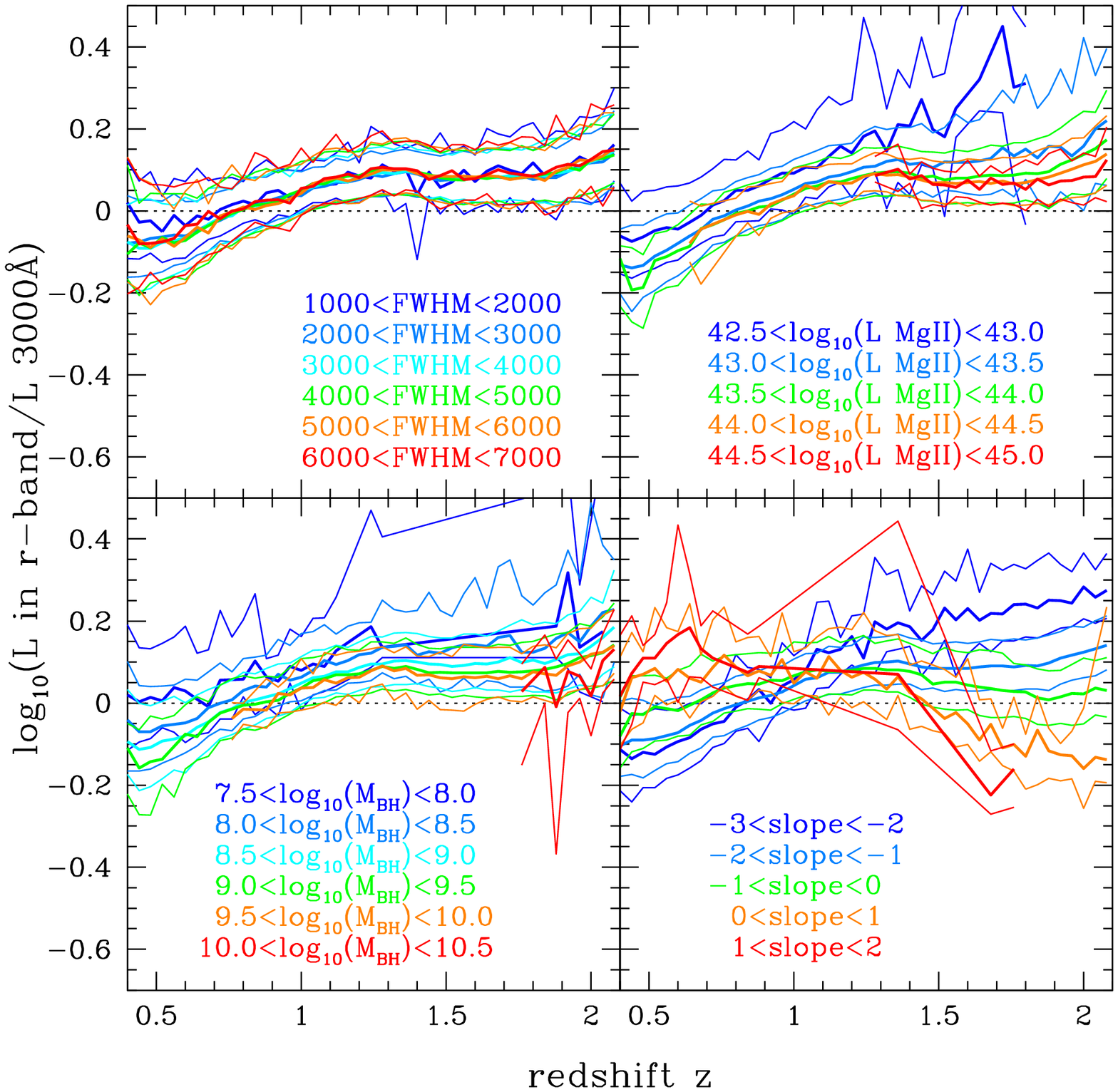}
\FigCap{Correlations between the measured median ratios of the $r$-band to
3000~\AA\ luminosity with the MgII emission line width and strength,
black hole mass, and the spectral slope. There is no obvious correlation
with the MgII FWHM (expressed here in km/s), but there is a weak
hint of anti-correlation with the MgII strength (in erg/s) and
the black hole mass (in \MS). There is, however, a strong
correlation of the ratios with the continuum slope, and changing with
redshift. We note that over 91\% of quasars have the slope in a range
$-2<{\rm slope}<0$ (green--light-blue). The AGNs with $-2<{\rm
slope}<-1$ and $-1<{\rm slope}<0$ have ratios different by at most
$1\sigma$ or 0.1~dex across the entire redshift range. While the ratios
for AGNs with $0<{\rm slope}<2$ (orange--red) do not follow the ones for
the main sample, they constitute only 2\% of the sample.}
\end{figure}

\Section{Summary}
In this paper, we have been interested in empirical conversions of
broad-band AGN magnitudes to the monochromatic luminosities that are
essential in calculating the bolometric AGN luminosity, central black hole
mass {\it via} the radius--luminosity relation, or simply the BLR
radius. Using the 105\,783 SDSS DR7 quasars, we calculate the ratios between
AGN luminosities as observed in common astronomical filters and their
monochromatic luminosity as a function of redshift. We provide a simple
prescription for using the broad-band magnitudes to calculate monochromatic
luminosity at 5100~\AA, 3000~\AA, and 1350~\AA.

AGNs do have different spectral shapes and different contributions from the
emission lines, hence the median values derived and provided in this paper
should rather serve as ``best estimates'' and not as ``measurements''.
Ratios for low redshift AGNs with $z<0.5$ may also be affected by AGN host
contamination, and hence are less reliable than the estimates at higher
redshifts. Galaxies are brightest in infrared, therefore the infrared
estimates (2MASS, WISE) should be used with even higher caution, as shown
in the discussed example above. We also study correlations of derived
conversions with the black hole mass, emission lines strength and width,
and the spectrum slope. Only the latter has a noticeable impact (of up to
$\approx0.1$~dex) on the derived conversions, but in the absence of spectrum
or when the spectrum is of low quality, this is a low price to pay for a
monochromatic luminosity estimate.

Since the majority of observations were taken at significantly different
epochs, we estimate that a fair fraction of the reported uncertainties is
not related to the intrinsic AGN properties, but simply due to variability.

Having the broad-band magnitudes converted to any or all the monochromatic
luminosities, it is straightforward to estimate the BLR radius using the
BLR-radius--luminosity relation (Kaspi \etal 2000, Bentz \etal 2009).
Transformations provided in this paper may also serve as tools in designing
future spectroscopic and/or photometric reverberation mapping campaigns,
similar to the one reported in Shen \etal (2015).

Bolometric luminosities ($L_{\rm bol}$) are an important diagnostic for AGN
studies, as they are necessary in calculating the Eddington ratio ($=L_{\rm
bol}/L_{\rm Edd}$, where $L_{\rm Edd}=1.26\times10^{38}(M_{\rm
BH}/\MS)$~erg/s), and what follows, the mass accretion rate (see \eg\break
Peterson 1997). Bolometric luminosities of AGNs can be calculated from
their monochromatic luminosities, where these fluxes are ``simply
multiplied'' by 9.26, 5.15, and 3.18 for 5100~\AA, 3000~\AA, and 1350~\AA\
(see discussion in Shen \etal 2011 and/or Richards \etal 2006b). Here, they
can be estimated by adding 0.97, 0.71, and 0.50 in Eq.(3), respectively.

We provide a simple online calculator of monochromatic and bolometric AGN
luminosities based on the analytic conversions from Eqs.(2--3) and
Tables~3 and 4:
\begin{center}
{\it http://www.astrouw.edu.pl/\textasciitilde simkoz/AGNcalc}
\end{center} 

\Acknow{This work has been supported by the Polish National Science Centre
grant No. 2014/15/B/ST9/00093. We thank Prof.~Christopher
S.~Kochanek for careful reading of the manuscript and many comments that
helped to improve it. We also thank Dr.~Kelly Denney for helpful
comments on the manuscript. This research has made use of the SIMBAD
database, operated at CDS, Strasbourg, France. This research has made
use of the NASA/IPAC Extragalactic Database (NED) which is operated by
the Jet Propulsion Laboratory, California Institute of Technology, under
contract with the National Aeronautics and Space Administration.}

\end{document}